\documentclass{article}

\usepackage{PRIMEarxiv}

\usepackage[utf8]{inputenc} 
\usepackage{amsmath}
\usepackage{tabularx}
\usepackage[T1]{fontenc}    
\usepackage{hyperref}       
\usepackage{url}            
\usepackage{booktabs}       
\usepackage{amsfonts}       
\usepackage{nicefrac}       
\usepackage{microtype}      
\usepackage{lipsum}
\usepackage{fancyhdr}       
\usepackage{graphicx}       
\graphicspath{{media/}}     

\usepackage{multirow}
\usepackage{makecell}
\title{Free-Space Optical Spiking Neural Network}

\author{
  Reyhane Ahmadi \\
Department of Computer Engineering\\
	Sharif University of Technology\\
	Tehran, Iran, 11155-4365 \\
\texttt{reyhaneahmadi1234@yahoo.com} \\
   \And
  Amirreza Ahmadnejad \\
Department of Electrical Engineering \\
	Sharif University of Technology\\
	Tehran, Iran, 11155-4365 \\
  \texttt{amirreza.ahmadnejad@sharif.edu } \\
     \And
  Somayyeh Koohi\\
  Department of Computer Engineering \\
	Sharif University of Technology\\
	Tehran, Iran, 11155-4365 \\
  \texttt{koohi@sharif.edu } \\
}

\begin{document}
\maketitle

\begin{abstract} 
Neuromorphic engineering has emerged as a promising avenue for developing brain-inspired computational systems. However, conventional electronic AI-based processors often encounter challenges related to processing speed and thermal dissipation. As an alternative, optical implementations of such processors have been proposed, capitalizing on the intrinsic information-processing capabilities of light. Within the realm of optical neuromorphic engineering, various optical neural networks (ONNs) have been explored. Among these, Spiking Neural Networks (SNNs) have exhibited notable success in emulating the computational principles of the human brain.
Nevertheless, the integration of optical SNN processors has presented formidable obstacles, mainly when dealing with the computational demands of large datasets. In response to these challenges, we introduce a pioneering concept: the Free-space Optical deep Spiking Convolutional Neural Network (OSCNN). This novel approach draws inspiration from computational models of the human eye. We have meticulously designed various optical components within the OSCNN to tackle object detection tasks across prominent benchmark datasets, including MNIST, ETH 80, and Caltech. Our results demonstrate promising performance with minimal latency and power consumption compared to their electronic ONN counterparts. Additionally, we conducted several pertinent simulations, such as optical intensity-to-latency conversion and synchronization. Of particular significance is the evaluation of the feature extraction layer, employing a Gabor filter bank, which stands to impact the practical deployment of diverse ONN architectures significantly.

\end{abstract}

\section{Introduction}
The human brain represents a profoundly intricate and remarkable biological entity. The endeavor to engineer a computational processor possessing commensurate attributes in power, precision, integration, and speed has perennially constituted a paramount aspiration for processor designers. Neuromorphic Engineering (NE) \cite{RN1} stands as a foundational paradigm facilitating the realization of such processors, primarily through the incorporation of neural network architectures (NNs). Despite the notable achievements resulting from this approach \cite{RN2,RN3}, the central challenge in processor design endures as the demand for processing voluminous datasets continues to burgeon.
To address this persistent challenge, optical Neuromorphic Engineering has emerged as a novel and innovative domain. Optical Neuromorphic Engineering exploits the distinctive attributes of light, which include its exceptional propagation speed and the extended degrees of freedom it affords in comparison to electrons, encompassing characteristics like path, frequency, phase, polarization, and mode. Furthermore, optical systems manifest reduced loss, rendering them remarkably compelling for the design and construction of Optical Neural Networks (ONNs).

The pursuit of processor miniaturization has perennially remained a core objective in the field of processor design \cite{RN4}. This pursuit has extended to the domain of Optical Neural Networks (ONNs) \cite{RN5}. Nevertheless, the efficient processing of substantial datasets at elevated speeds presents a considerable challenge within this sphere. In response to this challenge, recent research endeavors have refocused on developing processors harnessing the capabilities of optical free space (OFS) devices \cite{RN6,RN7,RN8,RN9,RN27}.

Spiking Neural Networks (SNNs), constituting a class of neural networks that emulate the structural and functional aspects of the human brain, have garnered significant attention in this context. Many optical models have been proposed for implementing SNNs; however, these models have been primarily integrated. Several intricate designs featuring components such as Vertical-Cavity Surface-Emitting Lasers (VCSELs) \cite{RN10}, micro ring resonators \cite{RN11}, and phase-change materials \cite{RN12} have been suggested. Nonetheless, these designs prove ill-suited for the demanding task of high-volume data processing, thereby underscoring the formidable challenges encountered within the realm of optical Neuromorphic Engineering.

To develop an Optical Free Space (OFS) model for Spiking Neural Networks (SNNs), it is imperative to establish precise mathematical models for each constituent component. Remarkably successful models rooted in neuroscience \cite{RN13,RN14} have been devised, emulating the structural attributes responsible for object detection within the human eye. We aim to draw upon these well-established models as a source of inspiration for designing OFS components, thereby facilitating the simulation of the Free-Space Optical deep Spiking Convolutional Neural Network (OSCNN). To the best of our knowledge, OSCNN marks the inaugural foray into the realm of OFS modeling for SNNs, encompassing critical elements such as Gabor filters for feature extraction, intensity-to-delay conversion, synchronization mechanisms, convolution layers, Max-pooling procedures, and a classification framework.

To execute the intensity-to-delay conversion, a dedicated module was introduced, employing a Spatial Light Modulator (SLM) after the feature extractor layer. Moreover, an optical synchronizer has been meticulously devised to address the temporal processing aspects inherent to optic signals. Ensuring that the time order of signals remains intact post-convolution, this synchronizer draws inspiration from the Free-Space Optical (FSO) delay line concept \cite{RN15}.

The performance of OSCNN was systematically evaluated across three distinct datasets: MNIST, Caltech, and ETH80. OSCNN demonstrated notable achievements, boasting significant performance metrics compared to electronic Neural Networks (NNs) and Optical Neural Networks (ONNs). Notably, Gabor filters were harnessed as feature extractors in the initial model layer, with evaluations conducted under both trained and fixed conditions. While alternative filters, such as Canny, Laplacian, and Sobel, were explored as feature extraction mechanisms, the most favorable outcomes for OSCNN were attained using Gabor filters. The results underscore the versatility of Gabor-form convolutional kernels, revealing their efficacy in image and time-series processing applications \cite{RN25,RN33}.

\section{OSCNN Model}

In delineating the architectural framework of the Optical Free Space Spiking Convolutional Neural Network (OSCNN), it is imperative to establish a comprehensive model for neurons within Spiking Neural Networks (SNNs, also known as integrate-and-fire models). SNNs fundamentally operate on the principles of spike-timing-dependent plasticity (STDP), a paradigm necessitating optical modeling to faithfully replicate its mechanisms. OSCNN adopts a specialized variation of computational neurons, acknowledging that forthcoming research will delve into developing a more precise neuron model. It is worth noting that as an alternative to STDP, backpropagation (BP) can be employed for training the OSCNN model, as described in \cite{RN16}. Subsequently, this section elaborates on the mathematical underpinnings of each module within OSCNN, along with their optical equivalents. The holistic structure of the OSCNN model is illustrated in Figure 1 for a comprehensive overview.

\begin{figure}[ht!]
\centering
\includegraphics[width=13cm]{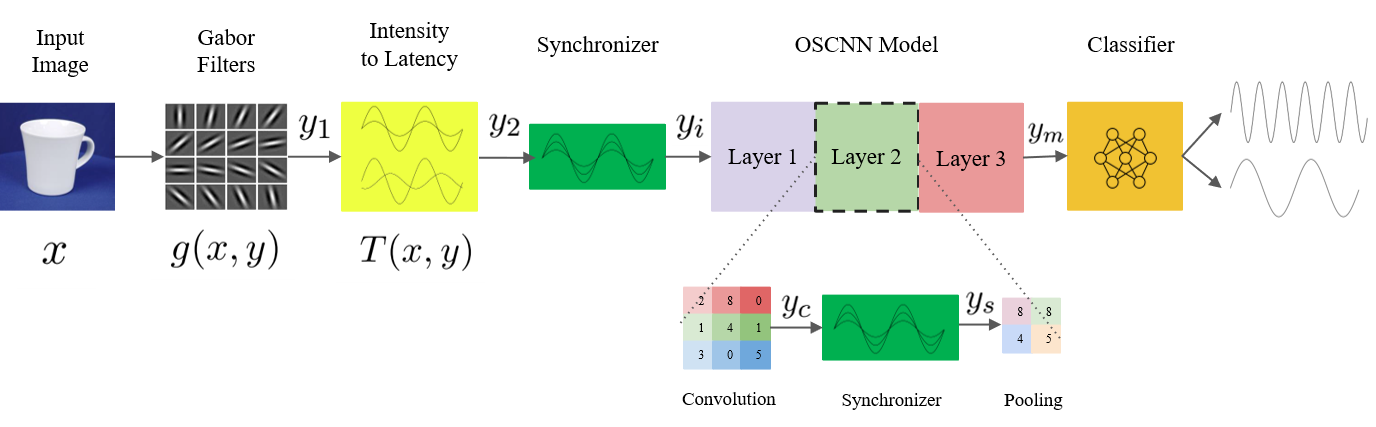}
\caption{Outline of OSCNN.}
\end{figure}

\subsection{Gabor Filters}

A Gabor filter serves as a bandpass filter meticulously designed to emulate the receptive field characteristics of neurons in the primary visual cortex of the human brain \cite{RN17}. This filter is uniquely characterized by two key parameters: the frequency of the sinusoidal wave and the width of the Gaussian envelope. In implementing feature extraction via convolution using the Gabor kernel filter, an optical 4f correlator emerges as a viable solution \cite{RN7,RN27}.

To execute the convolution of an image, denoted as $x$, with a Gabor filter represented by $g$, a Fourier domain approach is adopted. In this process, the Gabor filter and the input image are transformed into their respective Fourier domain representations. Subsequently, these transformed representations are multiplicatively combined within the Fourier domain. The Fourier transform of the Gabor filter is expressed as follows:
\begin{equation}
\label{deqn_ex1a}
\begin{split}
G(u, v) = e^{-\frac{1}{2}\left(\frac{\sigma_x^2(u-k_x)^2 + \sigma_y^2(v-k_y)^2}{\sigma_x^2\sigma_y^2}\right)}
\end{split}
\end{equation}

The Fourier transformation of the Gabor filter reveals itself as another Gabor function, characterized by specific frequency components, denoted as $k_x$ and $k_y$, alongside standard deviations represented by $\sigma_x$ and $\sigma_y$. Consequently, the mathematical model for the convolution of an image $X(u,v)$ with a Gabor filter through the utilization of an optical 4f correlator can be expressed in the following manner:
\begin{equation}
\label{deqn_ex1a}
\begin{split}
y_1 (u, v) = X(u, v) \cdot G(u, v)){\sigma_x^2\sigma_y^2}
\end{split}
\end{equation}

\subsection{Intensity to Phase Conversion}

The resultant features manifest as light patterns characterized by varying intensities. To faithfully represent the optical properties and mitigate signal losses, a crucial transformation process is invoked, converting the information modulated on intensity in the input image into phase information. This transformation is known as intensity-to-phase (latency) conversion and is effectuated through a Spatial Light Modulator (SLM).

Suppose $y_1(u,v)$ symbolizes the output derived from the preceding convolution module, associated with intensity values $I(u,v)$. In that case, the conversion operation can be articulated as $I(u,v) = I_0\cos^2(\phi(u,v))$, where $I_0$ designates the maximum intensity and $\phi(u,v)$ signifies the phase values. To effectuate this conversion, the SLM introduces a spatially varying phase shift commensurate with the intensity at each pixel. Mathematically, the SLM can be represented as a complex transmission function $T(u,v)$, satisfying the condition $|T(u,v)| \leq 1$. In most instances, $T(u,v)$ is expressed as $e^{i\phi(u,v)}$, where $\phi(u,v)$ corresponds to the desired phase shift at each pixel. Hence, the complex transmission function assumes the form of a complex exponential, featuring a phase term directly proportional to the desired phase shift. Consequently, the output stemming from the intensity-to-phase conversion module is succinctly represented as:

\begin{equation}
\label{deqn_ex1a}
\begin{split}
{y}_2(u,v) = y_1(u,v) \cdot T(u,v)
\end{split}
\end{equation}

\subsection{Optical Synchronizer}

Upon converting intensity to phase, it is imperative to address that each optical signal propagates at varying speeds, with the fastest signal being of utmost significance. To facilitate the concurrent processing of all signs and their combinations, it is essential to ensure their emission commences from a common temporal reference point. This task is conventionally achieved in integrated ONNs by employing a delay line structure \cite{RN2}. In contrast, this practice is less prevalent within Optical Free Space (OFS). In OFS, the synchronization is typically accomplished through a diffraction grating and a Spatial Light Modulator (SLM) instead of the traditional delay line or parallel mirrors \cite{RN9}.

A synchronizer is engineered by leveraging an SLM and a diffraction grating by the principles delineated in \cite{RN15}. The output field originating from the preceding module is $y_2(u,v)$, subjected to diffraction through a grating characterized by a specific pitch denoted as $\Lambda$, along with a designated diffraction angle, $\theta$. The output emerging from the grind embodies multiple diffraction orders, each encompassing a version of the input signal delayed by a distinct temporal offset. The electric field associated with the $m$th diffraction order is aptly described as $E_m(u,v)$, featuring a time delay of $\tau_m = m\Lambda\sin\theta/c$. Consequently, the electric field affiliated with the $m$th diffraction order can be succinctly articulated as:
\begin{equation}
\label{deqn_ex1a}
\begin{split}
E_m (u,v) = y_2 (u,v) e^{i2\pi m \Lambda sin\theta /c}
\end{split}
\end{equation}

After the diffraction grating, the individual diffraction orders undergo phase modulation through the utilization of an SLM, where the relative phases of the input signals are meticulously adjusted. The phase modulation introduced by the SLM can be mathematically represented by a complex-valued function $f(u,v)$. To reconstitute the electric fields associated with the diverse diffraction orders, a lens boasting a focal length denoted as $f$ is employed. Ultimately, the output signal is derived from the electric field situated at the focal point of the lens. This output signal can be expressed as (with $\mathrm{sinc}(x) = \frac{\sin(\pi x)}{\pi x}$):
\begin{equation}
\label{deqn_ex1a}
\begin{split}
y_i(u,v)=\sum_{m=-\infty}^{\infty} E_m (u,v) e^{i f(u,v)} sinc(\frac{m\Lambda sin\theta}{f})
\end{split}
\end{equation}

\subsection{Layers and Classifier}

Following the feature extraction phase, the optical signals must be amalgamated to capture the salient features embedded within the image. This synthesis is achieved via a 3-layer system comprising a convolution layer, a synchronizer, and a max-pooling module \cite{RN14}. The convolution layer effectively combines various optical signals, each subject to different learnable weights, while the max-pooling layer identifies and preserves the most crucial features. The necessity for a synchronizer arises from the typical propensity of the convolution operation to disarrange the temporal order of signals. The convolution operation is mathematically represented as a 4f correlator, characterized by trainable kernels \cite{RN7, RN27}. In parallel, the max-pooling function is realized by employing a 4f correlator, incorporating a saturable absorber (SA) as a nonlinearity, which can be conceptually likened to a multilayer neural network augmented with nonlinearity \cite{RN7,RN18}.

The optical emulation of a classifier is inspired by the approach delineated in \cite{RN7}. It consists of a single classifier utilizing an MNN with a non-linearity module, specifically the saturable absorber (SA). The comprehensive structural layout of the OSCNN model is prominently depicted in Figure 2.

\begin{figure}[ht!]
\centering
\includegraphics[width=13cm]{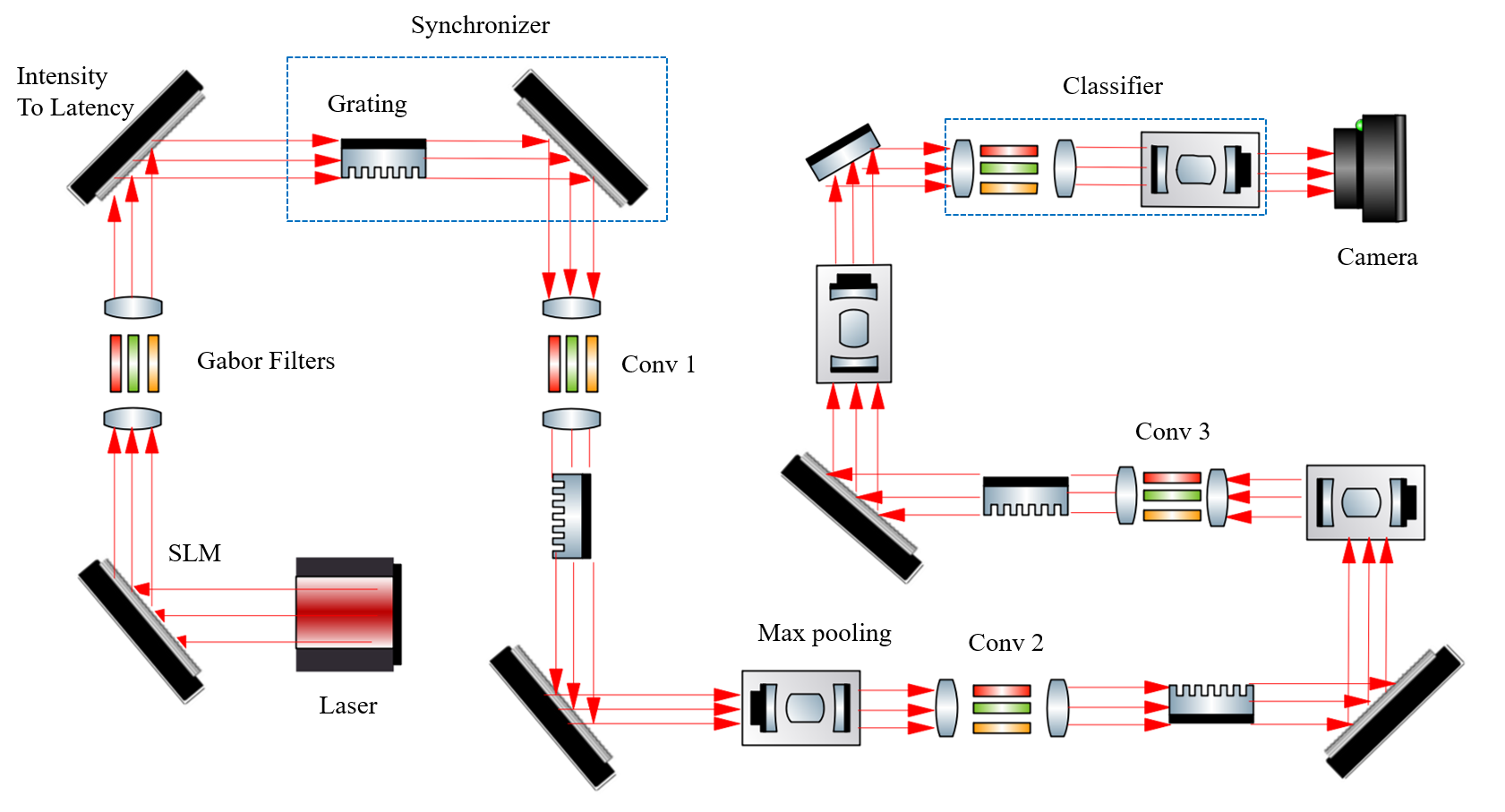}
\caption{The simulation of OSCNN involves loading data from a laser and SLM and simulating different optical free space components. }
\end{figure}

\section{Simulation Study}

This section delves into the comprehensive simulations conducted to evaluate the performance of the Free-Space Optical Spiking Convolutional Neural Network (OSCNN). The study extends to comparative analyses with other models, encompassing both electrical and optical domains, encompassing free-space and integrated approaches. A primary focus of the investigation lies in the in-depth analysis of the first-layer feature extractor kernels in both fixed and trainable configurations. Incorporating Gabor filters as convolutional kernels in the feature extractors of Convolutional Neural Networks (CNNs) is particularly emphasized, owing to its biological inspiration from the human brain and inherent properties. This is further juxtaposed with comparisons involving other well-recognized filters such as Sobel, Canny, and Laplacian. Furthermore, the impact of noise on the input image is methodically examined, and the temporal consumption of electrical or optical resources by the OSCNN is quantitatively measured.

The OSCNN model was subjected to rigorous training using the MNIST dataset, facilitated by a V100 Tesla GPU on the Google Colab platform. The cumulative duration for processing and training with the MNIST dataset amounted to 2 hours and 37 minutes, a timeframe that is markedly consistent with processing times associated with other electrical and optical models, as elucidated in \cite{RN7,RN8,RN19,RN27}. The training process was executed through backpropagation, functionally equivalent to the Spike-Timing-Dependent Plasticity (STDP) process described in \cite{RN16}.

In addition to the MNIST dataset, the model was subjected to rigorous training and testing on the ETH-80 dataset \cite{RN20} and the Caltech dataset \cite{RN21}, both of which are well-recognized benchmarks within the Spiking Neural Network (SNN) domain \cite{RN14}. The simulation of the OSCNN closely aligns with established methodologies applied in prior models, such as those documented in \cite{RN7,RN8,RN9,RN24,RN27}. Each optical module is rigorously formulated mathematically within these simulations and is referred to as a behavioral model.

\subsection{Gabor Filters}

The initial layer of the OSCNN is characterized by an array of Gabor filters, each possessing distinct spatial orientations and thicknesses. To provide a comprehensive comparative analysis of the feature extraction process, we benchmark the OSCNN against other Optical Neural Networks (ONNs). For instance, the Diffractive Deep Neural Network (D2NN) \cite{RN22} leverages light diffraction properties employing apertures designed through the Huygens principle. In the OSCNN, Gabor filters are deployed to extract the most salient features embedded within the images meticulously. To this end, various feature extraction approaches are examined, including fixed Gabor filters devoid of training, trainable Gabor filters, and established filters like Canny, Laplacian, and Sobel. These diverse feature extractors are employed with the MNIST, Caltech, and ETH-80 datasets. The ensuing influence of these filters on the output accuracy is meticulously assessed and is presented in Table 1.

\begin{table}[htbp]
\centering
\caption{OSCNN accuracy with different feature extractors}
\begin{tabular}{|c|c|c|c|}
\hline
Filter & MNIST & Caltech & ETH-80 \\
\hline
Fixed Gabor & $91.4$ & $88.3$ & $84.7$ \\
\hline
Trainable Gabor & $\bf{95.2}$ & $\bf{91.3}$& $\bf{89.7}$ \\
\hline
Canny & $89.2$ & $86.7$& $84.2$ \\
\hline
Laplacian & $90.4$ & $84.3$& $83.5$ \\
\hline
Sobel & $86.7$ & $82.1$& $79.4$ \\
\hline
\end{tabular}
  \label{tab:shape-functions}
\end{table}

The outcomes, as depicted in Table 1, reveal that the trainable Gabor filter, endowed with adaptable parameters about its filter length and central frequency, attains the highest performance among the filters tested. However, it is noteworthy that even the fixed Gabor filter consistently outperforms the alternative filters. This substantiates our assertion that Gabor filters can be regarded as reliable and effective feature extractors across various Optical Neural Networks (ONNs) and can be effectively deployed in the first layer as Convolutional Neural Network (CNN) kernels. For further visual representation, Figure 3 showcases the output images generated by applying the Gabor input filter to an image from the dataset, exemplified by image number 8.

\begin{figure}[ht!]
\centering
\includegraphics[width=12cm,height=3cm]{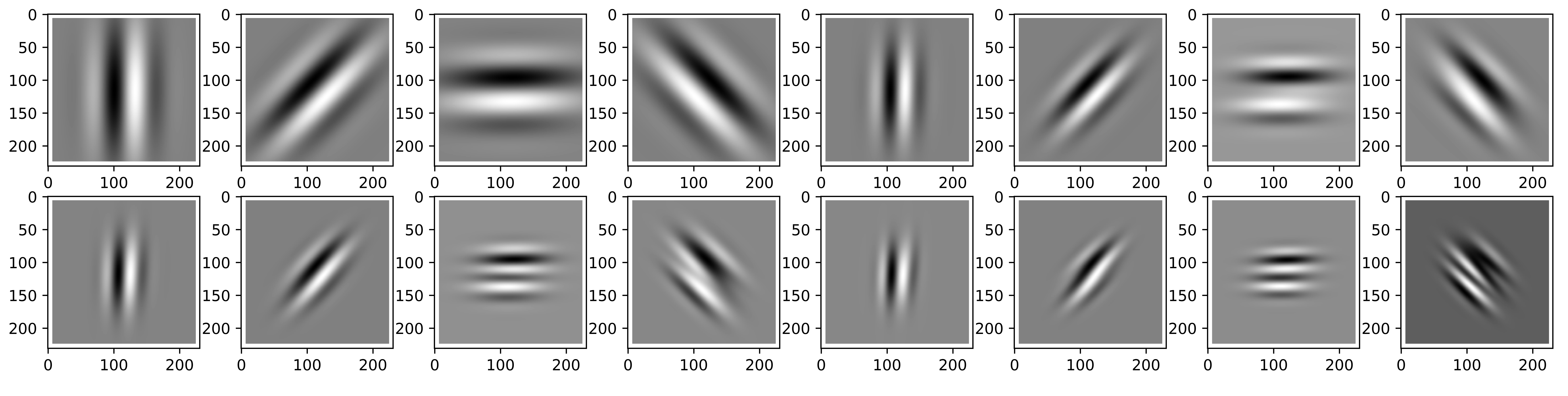}
\caption{Gabor filters are created as sine wave filters in four different directions, namely $0$, $\frac{\pi}{2}$, $\pi$, and $\frac{3\pi}{4}$ degrees. These filters have different thicknesses and function as edge detectors, resembling the simple cells found in the primary visual cortex \cite{RN17}. }
\end{figure}

This comprehensive analysis underscores the exceptional efficacy of Gabor filters as optimal models for feature extraction in the initial layer of Optical Neural Networks (ONNs). Consequently, Gabor filters stand as a highly recommended choice for the processing of diverse data types, extending their applicability to various domains, including image analysis and the handling of temporal signals, such as audio, as expounded in \cite{RN25}.

\subsection{Intensity to Phase Conversion}

Employing a Spatial Light Modulator (SLM) for intensity-to-phase conversion within OSCNN necessitates meticulous considerations to ensure optimal performance. These considerations encompass spatial resolution, phase modulation range, and the selection of the optimal operating wavelength. The spatial resolution of the SLM plays a pivotal role in defining the output spatial resolution of the OSCNN. Therefore, a high-resolution SLM is highly desirable for achieving superior spatial resolution. Furthermore, the phase modulation range exhibited by the SLM profoundly influences the dynamic range of the OSCNN, with a more extensive phase modulation range facilitating a broader dynamic range. The SLM's specifications must align optimally with the operating wavelength of the optical system employed within the OSCNN to ensure maximum efficiency and precision.

While the simulations affirm that the SLM is one of the most straightforward and practical choices for implementing Intensity Latency conversion, alternative methods are available, including Amplitude-Phase Modulation (APM) and holographic filtering. The selection among these methods should be contingent on the specific requirements of the application, encompassing factors such as speed and power consumption. To provide a comparative assessment of processing speed for a $28\times 28$ pixel image with Gabor filters, followed by intensity to latency conversion, the time required for the SLM stands at approximately $1 ms$ \cite{RN28}. In contrast, for APM, it is roughly $0.5 ms$ \cite{RN29}, and for holographic filtering, it approximates 1 ms \cite{RN31}. Therefore, the speed comparison can be briefly summarized as follows:
\begin{equation}
t_{APM} < t_{SLM} < t_{HF}
\end{equation}

Where $t_{HF}$ is the time required for intensity to latency conversion using holographic filtering.

In power consumption, it is essential to note that SLMs primarily leverage electro-optic effects, which typically demand power in the millijoules per second range \cite{RN30}. In contrast, Amplitude-Phase Modulation (APM) leans on nonlinear effects like Pockels or Kerr, which may escalate power consumption to the level of joules per second \cite{RN29}. Therefore, the comparison of power consumption can be concisely summarized as follows:
\begin{equation}
P_{SLM} = P_{HF} < P_{AMP}
\end{equation}

\subsection{Synchronizer}

The accurate simulation of a synchronizer hinges on a meticulous planning process, necessitating careful consideration of the diffraction angles and grating period. This planning is crucial to ensure that the diffraction orders are distinctly separated and that the resulting synchronized output field boasts a superior signal-to-noise ratio. The Spatial Light Modulator (SLM) plays a pivotal role in this endeavor by encoding the filter as a phase pattern. Precise design of the phase pattern on the SLM, in conjunction with careful spacing of the grating, enables the synchronization of distinct diffraction orders, permitting their simultaneous arrival and concurrent processing.

A diffraction grating, in this context, assumes the form of a periodic structure replete with regularly spaced lines or slits, adept at diffracting incident light into a specific pattern. When designing a diffraction grating, two paramount parameters must be meticulously considered: the grating period, denoted as $d$, and the number of lines, represented by $N$. To craft a grating tailored for a specific wavelength, the following formula proves instrumental:

\begin{equation}
d sin(\theta) = m\lambda 
\end{equation}

In our simulation of the OSCNN, we employ a red laser with a wavelength of $1550 nm$, and the separation between the SLM and the diffraction grating is $5 cm$. For this scenario, we consider diffraction of the incident light at an angle of $5 ^{\circ}$ while specifically utilizing the first diffraction order, $m = 1$. As such, referring to equations 8 , we calculate the grating period to be $11.8 \mu m$, and the corresponding diffraction angle is approximately $32.8 ^{\circ}$.

\subsection{Model Performance}

The evaluation of OSCNN's proficiency in object detection tasks encompassed its rigorous training and testing on three distinct datasets: MNIST, ETH 80, and Caltech. Subsequently, the results were meticulously juxtaposed against established electronic and optical models, both in the free-space and integrated domains. These comparative results are illustrated in Figure 4 and tabulated in Table 2.

Figure 4 offers a comparative overview, contrasting the performance of electronic implementations, specifically those of Alexnet and electrical models, with their OFS counterparts. It is noteworthy that SNNs, whether instantiated electronically or optically, consistently manifest superior object detection performance when compared to CNNs.

\begin{figure}[ht!]
\centering
\includegraphics[width=12cm]{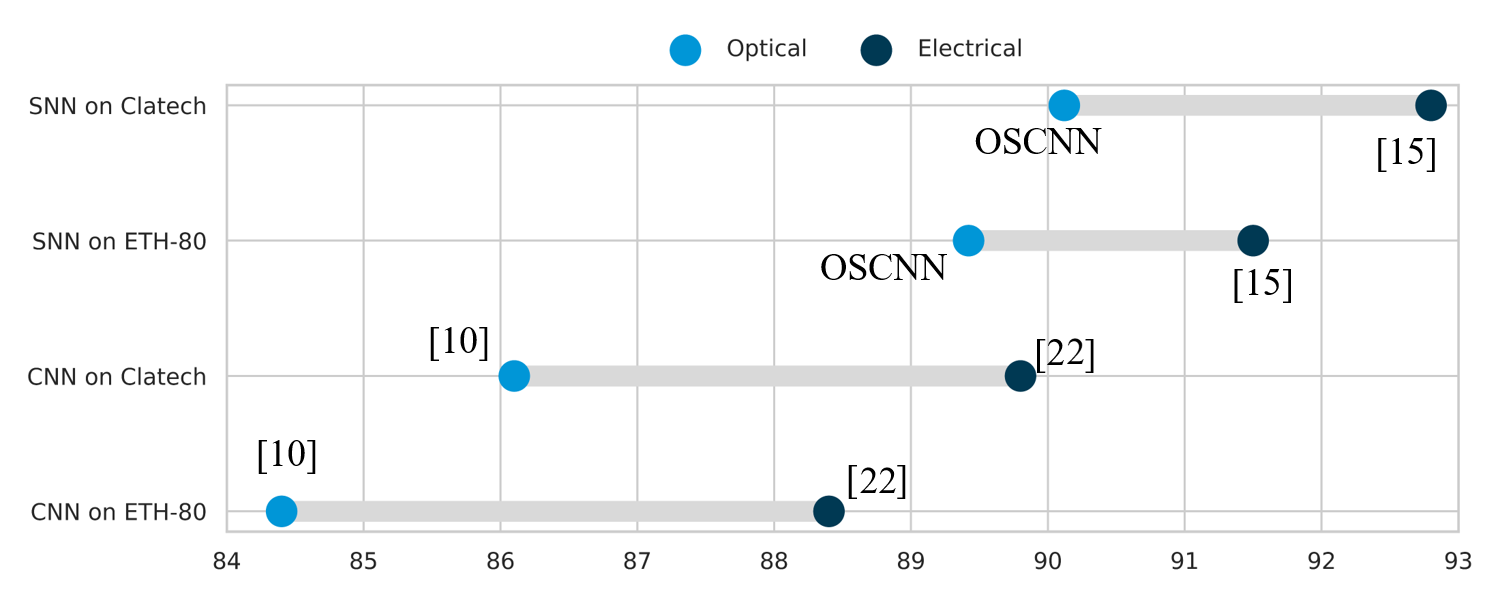}
\caption{Accuracy of different electrical and optical Neural Network on Caltech and ETH-80 data sets.}
\end{figure}

\begin{table*}
\centering
\caption{MNIST Object Detection Accuracy and Models}
\label{my-label}
\begin{tabular}{|c|c|c|c|c|}
\hline
\multicolumn{2}{|c|}{Implementation} & Model & MNIST Accuracy (\%) & Input Size \\
\hline
\multirow{3}{*}{Electrical} & Perception & MLP \cite{RN23} & 92.3 & $28\times28$ \\
\cline{2-5}
& Convolution & AlexNet \cite{RN19} & 98.3 & $28\times28$ \\
\cline{2-5}
& Spiking & Kheradpisheh et al. \cite{RN14} & 97.2 & $28\times28$ \\
\hline
\multirow{6}{*}{Optical} & \multirow{3}{*}{Perception} & Shen et al. \cite{RN3} & 95.4 & $3\times3$ \\
\cline{3-5}
& & D2NN \cite{RN20} & 99.1 & $3\times3$ \\
\cline{3-5}
& & Ryou et al. \cite{RN6} & 86.7 & $3\times3$ \\
\cline{2-5}
& \multirow{2}{*}{Convolution} & Bagherian et al. \cite{RN2} & 87.7 & $3\times3$ \\
\cline{3-5}
& & Sadeghzadeh et al. \cite{RN7} & 97.6 & $28\times28$ \\
\cline{2-5}
& \multirow{3}{*}{Spiking} & Feldmann et al. \cite{RN12} & 98.3 & $3\times3$ \\
\cline{3-5}
& & Xiang et al. \cite{RN11} & 89.9 & $3\times3$ \\
\cline{3-5}
& & OSCNN & 95.2 & $28\times28$ \\
\hline
\end{tabular}
\end{table*}

Table 2 serves as a comprehensive comparative analysis, shedding light on the performance of various electrical and optical models, all tested on the MNIST dataset. The findings unequivocally underscore that OSCNN exhibits commendable performance levels in the context of object detection tasks. Nevertheless, it is crucial to acknowledge a notable disparity between the implementations of NNs and ONNs, a distinction prominently illustrated in Figure 4 and elaborated upon in Table 2. In this context, it is essential to highlight that FSO implementations consistently demonstrate accelerated processing speeds when juxtaposed against their integrated counterparts. Furthermore, FSO implementations exhibit remarkable data-handling capabilities, extending to vast datasets, including biological data, a feat that remains presently unattainable for integrated ONNs.

\subsection{Noise Robustness}

The assessment of OSCNN extends to its resilience against various levels of input image noise. In our evaluation, we introduced white noise to the input images, thereby subjecting the model to different noise levels ranging from $5\%$ to $50\%$. Our findings regarding OSCNN's performance under varying noise conditions are vividly depicted in Figure 5. Furthermore, we subjected several other electrical and optical Neural Networks (NNs) to evaluations to gauge their robustness against noise. It is worth noting that up to a noise level of $15\%$, the model demonstrates a reasonable degree of accuracy. However, as noise levels surge beyond this threshold, accuracy experiences a steep decline, ultimately converging to the chance level of $40\%$.

\begin{figure}[ht!]
\centering
\includegraphics[width=8cm]{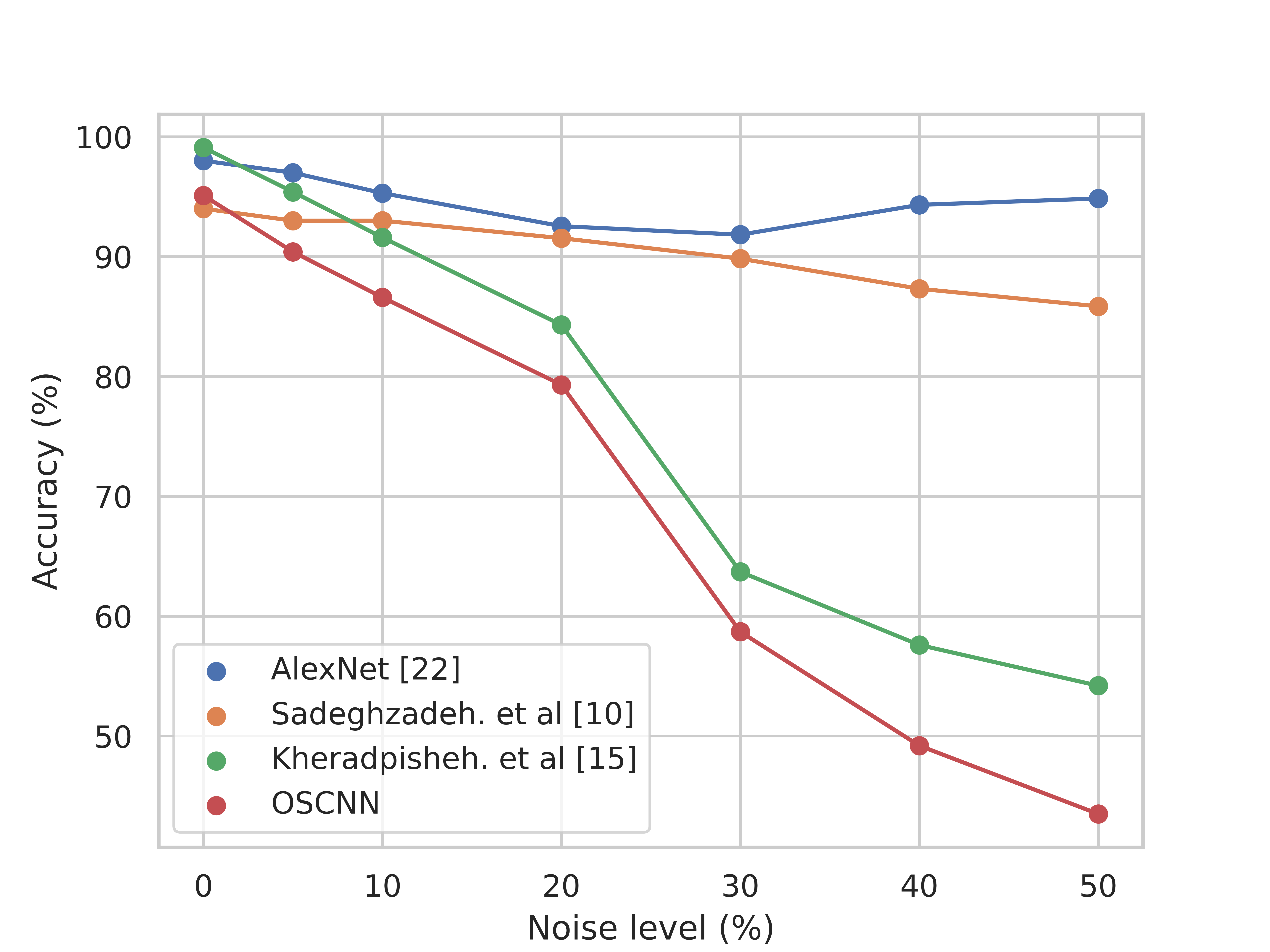}
\caption{Accuracy of different electrical and optical Neural Networks for various input noise levels.}
\end{figure}

Figure 4 offers a compelling visual representation of the comparative robustness exhibited by electrical CNNs and Optical formats, notably surpassing the robustness demonstrated by electrical SNNs and Optical ones. This observation prominently underscores one of the primary limitations inherent to SNNs, which is discernible not only in the electrical domain but also extends into the optical domain. Consequently, it becomes evident that further research endeavors are warranted to soothe and bolster the performance of SNNs, particularly in the context of robustness against noise.

\section{Speed Analysis}

Optical computing inherently harbors several compelling capabilities. The paramount advantage inherent to this optical implementation is its exceptional capacity for accelerated processing, distinctly outpacing its electrical counterparts. To provide a comprehensive insight into the relative speed of OSCNN, an intricate assessment of latency can be explained through the following estimation:

\begin{equation}
\begin{aligned}
t_{latency} & = t_{source}+t_{Gabor filters}+t_{PM}+t_{Sync}\\
& +t_{Conv1}+t_{Sync}+t_{Maxpooling}\\
& +t_{Conv2}+t_{Sync}+t_{Maxpooling}\\
& +t_{Conv3}+t_{Sync}+t_{Maxpooling}\\
& +t_{classifier}+t_{camera}+t_{transfer data}
\end{aligned}
\end{equation}

In this comprehensive analysis, the estimation of OSCNN's latency is meticulously deconstructed, considering various pertinent factors. To begin, $t_{source}$ signifies the modulation delay associated with input images. With the consideration of Spatial Light Modulators (SLMs) featuring a 1 kHz switching frequency, $t_{source}$ is approximated at $1 ms$ \cite{RN28}. Moving forward, when addressing $t_{Gabor filters}$, which fundamentally represents a convolutional layer employing Gabor kernels, the estimated processing time is approximately $5 ps$, a minuscule interval that can be safely disregarded. As for converting intensity into latency, a process executed by $t_{PM}$, the approximate duration allocated to this operation is also pegged at around $1 ms$ \cite{RN28}.

The synchronization system encompasses both a grating layer and an SLM. The dissonant layer's processing speed is approximately $1 ms$ for a $28 \times 28$ image \cite{RN15}, and the cumulative time required for processing the synchronization block is approximately $2 ms$. Subsequently, we arrive at $t_{Conv1}$ and $t_{Maxpooling}$, representing the optical propagation delays within the convolution and pooling layers. When employing 4f optical correlators, the processing delay for each 4f layer is approximately $5 ps$, equivalently, $5 ps$ for the pooling layers \cite{RN27}, durations that can be conveniently deemed negligible.

The classification module, integral to the system, comprises a 4F multiplier and a nonlinear Saturable Absorber (SA) unit. The delay attributed to the nonlinear unit is approximately $25 ns$ \cite{RN27}, while the classification component exhibits a delay of roughly $25 ms$. In this context, $t_{camera}$ represents the time required for photodetectors to capture and convert output images into electrical data. A high-speed commercial camera, capturing images at a rate of 2500 frames per second \cite{RN7}, is characterized by an estimated latency of $0.4 ms$. Lastly, the communication interface introduces delays when transmitting the camera's output data to a computer, as reflected by $t_{transfer data}$. By leveraging USB 3.1 Gen2 at a data rate of 10 Gbit/s and processing a 50 kB image, $t_{transfer data}$ is quantified at $0.04 ms$. Consequently, the cumulative delay attributed to OSCNN is reasonably estimated at approximately $2.44 ms$.

\section{Power Consumption Analysis}

The research outlined in \cite{RN27} sheds valuable light on the power dynamics within the OSCNN architecture. It delineates that Convolution, nonlinearity, and pooling operations are characterized by minimal energy consumption. Consequently, it follows that the principal contributor to energy consumption within the system lies in the domain of signal transmission \cite{RN27}. To contextualize these findings, it's essential to consider that each pixel's capture entails a power demand of 1W, as noted in \cite{RN7}:
\begin{equation}
\label{deqn_ex1a}
P_{optical}=\frac{n^2\times n_{kernel}}{\eta \times t^p} 
\end{equation}

The unit of power in this context is expressed in micro-watts. Within the equation, $n^2$ denotes the total number of pixels per 4f correlator system, while $p$ signifies the number of optical elements traversing the optical path. The variable $n_{kernel}$ is an arbitrary value denoting the number of kernels utilized by the convolutional layer. The parameter $t$ represents the fraction of incident power received by each optical element. Additionally, $\eta$ embodies the source efficiency. Conversely, when endeavoring to calculate the electric power consumption, careful consideration must be extended to the following factors, as expounded in \cite{RN27}:

\begin{equation}
\label{deqn_ex1a}
P_{electrical}=\beta \times n^2 \times k^2 \times n_{kernel}\times P_{switch}
\end{equation}

In the equation presented, the variable $\beta$ is determined by the architectural characteristics of the program, with $k$ signifying the kernel size and $P_{switch}$ representing the energy consumed by each operation. It's imperative to acknowledge that as the kernel size expands, electronic components invariably escalate their power consumption. Conversely, the optical implementation of convolutional layers showcases a distinct trend: as the kernel size diminishes, the power consumption concurrently decreases. This divergence in behavior highlights a remarkable aspect of optical implementations—namely, their capacity for significant power savings, particularly in scenarios featuring large kernel sizes, in contrast to their electrical counterparts.

It's also instructive to juxtapose these power considerations with the energy demands of the human brain during image processing. As evidenced by research in \cite{RN29}, the aggregate power consumed by the human brain for image processing hovers around 0.2 watts, accomplished within a processing timeframe of approximately ten milliseconds. In light of these findings, it becomes apparent that the optical neuromorphic approach, encompassing the optical creation of DNNs and SNNs, holds the potential to propel us a significant step closer to the realization of a processor endowed with computational prowess akin to the human brain, marked by its remarkable power efficiency.

\section{Conclusion}

This article marks a significant milestone by introducing the pioneering Optical Deep Spiking Convolutional Neural Network model operating in free space. Drawing inspiration from the computational model of the human eye, this model excels in detecting patterns with commendable accuracy, processing speed, and power efficiency. An essential revelation from this study is the suggestion that, in the realm of optical neural networks tasked with image processing, the initial layer can be effectively realized by deploying Gabor filters, thereby revolutionizing the approach to feature extraction. Moreover, the remaining optical free-space components, encompassing the Intensity-to-Delay conversion and the Synchronizer, were meticulously designed by harnessing the potential of readily available optical components. It's important to note that this optical model refrains from delving into the intricacies of biological neuron modeling and Spike-Timing-Dependent Plasticity (STDP) training. These aspects are deliberately deferred to future research endeavors, where the focus would entail designing a dedicated structure based on resonators or topological photonics. The objective is to authentically simulate the precise neuron model and subsequently introduce an array structure to usher in novel designs for convolutional layers, akin to the advancements witnessed in metasurface technology. Additionally, the study relies on a well-established behavioral model instead of a numerical analysis of electromagnetic fields. By introducing optical free space Deep Spiking Convolutional Neural Network models, a significant stride is taken toward the realization of high-powered, high-speed processors inspired by the human brain. This endeavor propels us closer to the development of an artificial brain, manifesting itself in an optical form endowed with formidable computational capabilities.

\section{Acknowledgments}

We acknowledge support from Dr. Zahra Kavehvash (Electrical Engineering Department of Sharif University of Technology) and Dr. Alireza Ejlali (Computer Engineering Department of Sharif University of Technology). We also would like to express our gratitude to OpenAI for the invaluable assistance provided by ChatGPT \cite{RN32} during the revision and refinement of this paper.

\bibliographystyle{unsrt}
\bibliography{sample}

\end{document}